\documentclass[a4paper]{mem}
\usepackage{natbib}
\usepackage{graphicx}
\begin{document}
   \title{Mid-infrared selection of Quasar-2s in Spitzer's First Look Survey
}

   \author{Mark Lacy \inst{1}, 
     Gabriela Canalizo \inst{2},\\ 
      Steve Rawlings \inst{3}, \\
      Anna Sajina \inst{4}\\
     Lisa Storrie-Lombardi, Lee Armus, Francine R.\ Marleau, Adam
Muzzin \inst{1}\fnmsep
}

   \offprints{Mark Lacy}
\mail{MS220-6, 1200 E. California Boulevard, Pasadena, CA~91125, USA }

   \institute{Spitzer Science Center, California Institute of Technology,
MS220-6, 1200 E. California Boulevard, Pasadena, CA~91125, USA \email{mlacy@ipac.caltech.edu}\\ 
              \and University of California, Riverside, CA, USA\\
              \and Astrophysics, University of Oxford, UK\\
              \and Astronomy, University of British Columbia, Canada}

   \abstract{We present early results from the spectroscopic follow-up of a 
sample of candidate obscured AGN selected in the mid-infrared from the 
Spitzer First Look Survey. Our selection allows a direct comparison of the 
numbers of obscured and unobscured AGN at a given luminosity for the first 
time, and shows that the ratio of obscured to unobscured AGN at 
infrared luminosities corresponding to low luminosity quasars is 
$\approx 1:1$ at $z\sim 0.5$. Most of our optically-faint candidate obscured 
AGN have the high-ionization, narrow-line
spectra expected from type-2 AGN. A composite spectrum shows evidence for 
Balmer absorption lines, indicating recent star-formation activity in the 
host galaxies. There is tentative evidence for a decrease in the obscured AGN
fraction with increasing AGN luminosity.

   \keywords{galaxies:active -- infrared radiation -- galaxies:distances and redshifts }
   }
   \authorrunning{Lacy et al.}
   \titlerunning{Mid-infrared selection of quasar-2s}
   \maketitle
%

\section{Introduction}

   \begin{figure*}
   \centering
   \resizebox{\hsize}{!}{\rotatebox[]{0}{\includegraphics{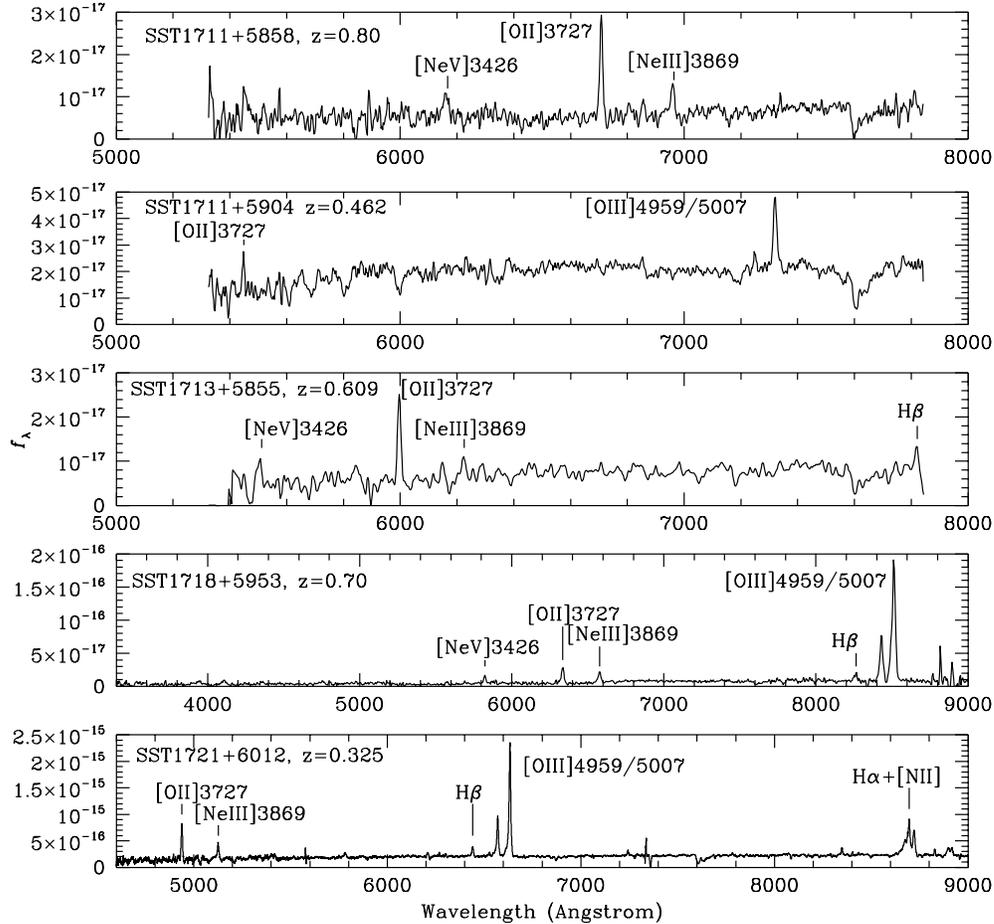}}}
   \caption{Optical spectra of the six most infrared-luminous of our 
objects to show the high-ionization, narrow-line spectra associated with 
type-2 AGN.}
              \label{FigGam}%
    \end{figure*}

The question of how many quasars are obscured by dust is still to be
resolved. Despite some success with samples selected in the hard
X-ray (e.g.\ Norman et al.\ 2002), in the near-infrared 
from 2MASS (Cutri et al. 2002, Glikman
et al.\ 2004) and from the Sloan Digital Sky Survey (SDSS) 
(Zakamska et al.\ 2004), there is still no
consensus on the number densities and luminosity distribution of the
hidden quasar population, particularly at high (quasar-like) luminosities.
There are a number of reasons for this. In the X-ray, it is still unclear
what fraction of high luminosity AGN might be Compton thick. 2MASS
selection suffers from not being able to select the most
heavily-extincted objects, and the SDSS selection on the basis of narrow
lines has potentially complex selection biasses. All these techniques
suffer from much uncertainty in relating the number densities of
quasars to those of quasar-2s when the two types of object are selected in
quite different ways. 

The fraction of hidden quasars is, however, 
important to establish if we are to
relate the mass density in black holes today to the accretion onto black
holes in the past (e.g. Yu \& Tremaine 2002). In particular, any
explanation for the close relationship between galaxy bulge velocity
dispersions and central black hole masses (e.g.\ Gebhardt et al.\ 2000)
involves an understanding of how black hole mass accretion varies with
epoch and host galaxy mass in both obscured and unobscured AGN. 

In Lacy et al.\ (2004; hereafter Paper 1) 
we presented a technique for selecting obscured 
AGN using only mid-infrared colours. The advantage of this technique
is that type-1 and type-2 AGN can be selected using the same criteria, 
removing the uncertainty involved when type-1 and type-2 objects are 
selected in different ways.
The effectiveness of this technique for selecting
AGN in {\em Spitzer} surveys has been confirmed by Stern et al.\ (2004) and 
Hatziminaoglou et al.\ (2004). In table 1 of Paper 1 
we presented a list of candidate obscured AGN. In this paper we present
early results from the spectroscopic follow-up of these objects.

\section{Optical spectroscopy}

Optical spectra were taken of twelve of the obscured AGN in Paper 1, 
concentrating on 
the fainter identifications most likely to be quasar-2s. 
The results are shown in Table 1 and Fig.\ 1. Most objects have the 
high-ionization, narrow-line spectra expected for quasar-2s, with several
either containing the [Ne{\sc v}]3426 emission line, or having sufficient
emission lines to place them firmly amongst the AGN in the diagnostic 
diagram of Kauffmann et al.\ (2003).

\begin{table*}
\caption{Results of optical spectroscopy}
\begin{tabular}{lcccl}
Object & $z$ & $S_{8\mu m}^{*}$&lg($L_{5\mu m}$ & Nature of optical spectrum\\
       &     & (mJy)           &   (WHz$^{-1}$))& \\ \hline
SSTXFLS J171106.8+590436&0.462&1.38&23.71&  high-ionization, narrow lines \\
SSTXFLS J171115.2+594906&0.587&5.09&24.30&  starburst spectrum\\
SSTXFLS J171147.4+585839&0.800&1.83&24.30&  high-ionization, narrow lines\\
SSTXFLS J171313.9+603146&0.105&4.65&22.90&  high-ionization, narrow lines\\
SSTXFLS J171324.3+585549&0.609&1.30&23.85&  high-ionization, narrow lines\\
SSTXFLS J171804.6+602705&0.43?&1.18&23.71&  single narrow emission line\\
SSTXFLS J171831.6+595317&0.700&1.22&24.27&  high-ionization, narrow lines\\
SSTXFLS J171930.9+594751&0.358&1.57&23.56& high-ionization, narrow lines\\ 
SSTXFLS J172050.4+591511& ?   &3.63&-& featureless red continuum\\
SSTXFLS J172123.1+601214&0.325&3.71&23.89&high-ionization, narrow lines\\
SSTXFLS J172328.4+592947&1.34?&1.69&25.12& single broad emission line\\
SSTXFLS J172458.3+591545&0.494&1.18&23.85&high-ionization, narrow lines\\
\end{tabular}

\noindent
$^{*}$ these values supercede those in Paper 1.
\end{table*}
   \begin{figure}
   \centering
   \includegraphics[width=2.2in]{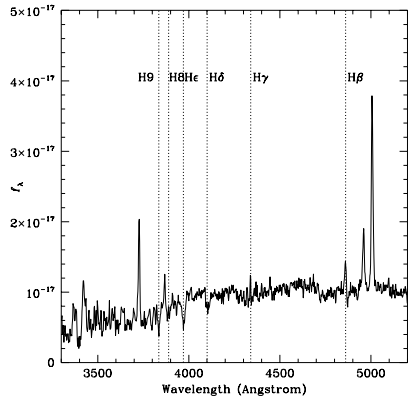}
      \caption{Composite spectrum showing Balmer absorption features in the host galaxies.}     
         \label{FigVibStab}
   \end{figure}

\section{Luminosities of the obscured AGN}
Using the redshifts, we can estimate
the luminosities of our obscured AGN. The mean ratio between the $B$-band and 
5$\mu$m luminosity density for quasars is about 10 (Elvis et al.\
1994), so a type-1 AGN on the quasar-Seyfert boundary (with $M_{B}=-22.3$
in our assumed cosmology of $H_0=70{\rm kms^{-1}Mpc^{-1}}$, 
$\Omega_{\rm M}=0.3, \, \Omega_{\Lambda}=0.7$)
would have a 5$\mu$m luminosity of 
${\rm lg}(L_{5\mu m}/{\rm WHz^{-1}})\approx 23.6$. 
Thus most of our spectroscopic targets have mid-IR luminosities corresponding
to low-to-moderate luminosity quasars (Table 1).
 
\section{The host galaxies}

Fig.\ 2 shows a composite spectrum made from five of the quasar-2, spectra 
(all taken with the Double Spectrograph of the Palomar 200$^{''}$). Balmer
absorption is clearly visible, indicating a significant A-star population, 
and thus star formation within the past $\sim 10^{8}$yr.

\section{Discussion}

Our spectroscopy has allowed us to confirm the result of 
Paper 1, namely, that 
at mid-IR luminosities corresponding to low-luminosity quasars, we find 
an $\approx 1:1$ ratio of type-2 to type-1 AGN. 
Most of our mid-IR selected obscured 
AGN seem to have unobscured narrow-line regions. Two
of our objects are not classifiable as AGN based on their optical spectra. 
One shows no emission lines, and the other has a starburst-like spectrum. 
Both these objects have extremely red mid-IR spectra, so may represent either
very deeply buried AGN with no emission able to escape to photoionize the 
narrow-line region, or starburst galaxies with extremely warm 
dust. Combining our spectroscopic redshifts 
with the photometric redshifts for the remainder of the sample of Paper 1 
shows that the median
redshift of our obscured AGN (0.46) is lower than that of
the unobscured population (0.69). 
Whether this is real, or a consequence of 
redshift-dependent selection effects remains to be determined. 
If real, it would be consistent with a luminosity-dependent 
obscured AGN fraction, and with the peak of obscured AGN activity
being at lower redshift, as determined from X-ray surveys (e.g.\ Cowie
et al.\ 2003). Urry \& Triester (2004), whose multiwaveband sample of AGN 
consists of  predominately lower-luminosity objects, 
find a 3:1 ratio of type-2 to type-1 AGN,
pointing to a luminosity-dependence of the obscured AGN fraction.
However, we expect
the effectiveness of mid-IR selection of AGN to decrease with
redshift as the host galaxy light becomes redshifted into the IRAC bands.
Modelling of mid-IR SEDs by Sajina, Lacy \& Scott (2004), 
suggests that we
should still be able to pick out obscured AGN up to $z\sim 2$ using the 
IRAC bands alone, however, the fact that our highest redshift AGN at 
$z=1.34$ shows a broad line in the optical suggests that
we may be losing a significant fraction of the more 
obscured objects at $z\sim 1$. Further
progress may depend on being able to include
MIPS 24$\mu$m fluxes in the AGN selection criterion.

\begin{acknowledgements}

We thank Alejo Martinez Sansigre for assitance with the WHT observations.
This work is based on observations made with the 
{\em Spitzer Space Telescope},
which is operated by the Jet Propulsion Laboratory (JPL), California Institute
of Technology under NASA contract 1407. Support for this work was provided
by NASA through JPL.

\end{acknowledgements}

\bibliographystyle{aa}

\begin{thebibliography}{}

\bibitem[]{} Cowie, L.L., et al.\ 2003, ApJ, 584, L57
Cutri, R., et al., 2002, in AGN Surveys, IAU colloquium 184, ASP conf ser 284, p.127 
\bibitem[]{} Elvis, M.\ et al.\ 1994, ApJS, 95, 1
\bibitem[]{} Gebhardt, K.\ et al.\ 2000, ApJ, 539, L13
\bibitem[]{} Glikman, E., et al.\ 2004, ApJ, 607, 60
\bibitem[]{} Hatziminaoglou, E.\ et al.\ 2004, AJ, in press (astro-ph/0410620)
\bibitem[]{} Kauffmann, G., et al., 2003, MNRAS, 346, 1055
\bibitem[]{} Lacy, M.\ et al., 2004, ApJS, 154, 156 
\bibitem[]{} Norman, C., et al., 2002, ApJ, 571, 218
\bibitem[]{} Sajina, A., Lacy, M.\ \& Scott, D., 2004, ApJ, in press (astro-ph/0409597)
\bibitem[]{} Stern, D.\ et al.\ 2004, ApJL, submitted (astro-ph/0410523)
\bibitem[]{} Urry, C.M.\ \& Treister, E.\ 2004, to appear in Growing Black
Holes, ed., A. Merloni, A. Nayakshin, R. Sunyaev (Springer-Verlag) (astro-ph/0409603)
\bibitem[]{} Yu, Q.\ \& Tremaine, S.\ 2002, MNRAS, 335, 965
\bibitem[]{} Zakamska, N.\ et al., 2004, AJ, 128, 1002 
\end{thebibliography}

\end{document}